\documentclass[cameraready]{Interspeech}

\title{GC-LoRA: Gated Convolutional LoRA for Parameter-Efficient Acoustic Adaptation}

\author[affiliation={1}, orcid=0009-0005-9726-6597]{Natarajan Balaji}{Shankar}
\author[affiliation={1}, orcid=0009-0006-3119-6159]{Zilai}{Wang}
\author[affiliation={1}, orcid=0009-0005-2488-3770]{Kaiyuan}{Zhang}
\author[affiliation={1}, orcid=0009-0009-5841-6846]{Mohan}{Shi}
\author[affiliation={1}, orcid=0009-0004-1406-503X]{Abeer}{Alwan}

\address{
    $^1$ University of California, Los Angeles, USA
}

\email{ \{balaji1312,zilaiwang2001,kaiyuanzhang,shimohan\}@ucla.edu, alwan@ee.ucla.edu}


\usepackage{comment}

\usepackage{multirow}

\begin{document}

\maketitle


\begin{abstract}
Transformer-based Speech Foundation Models excel in most Automatic Speech Recognition tasks but often suffer performance degradation when applied to domains with mismatched acoustic characteristics. While Parameter Efficient Fine-Tuning (PEFT) methods, such as Low-Rank Adaptation (LoRA), adjust global attention, they lack the local context modeling crucial for capturing domain-specific variations. We propose GC-LoRA, a novel adapter architecture that injects Conformer-style local convolutional processing into pretrained Transformer encoders. By integrating a lightweight adapter to encoder attention output projections, our method efficiently captures local acoustic dependencies without disrupting pretrained global representations. Experiments across diverse datasets (acoustically-degraded, bandlimited, dialectal, child) demonstrate the efficacy of our approach, achieving Word Error Rate (WER) reductions of up to 10.9\% compared to baselines while adding minimal trainable parameters.
\end{abstract}

\section{Introduction}
\label{sec:introduction}

Recent advances in Automatic Speech Recognition (ASR) have been driven by Speech Foundation Models (SFMs) \cite{chen2022wavlm,hsu2021hubert, canary, owsm}. SFMs trained on massive datasets, such as Whisper~\cite{radford2023robust}, exhibit strong zero-shot robustness and establish state-of-the-art performance on standard benchmarks. However, performance often degrades in target domains that diverge from pretraining data distributions, where data scarcity and domain mismatch remain major challenges \cite{chang2024self}. This vulnerability is particularly evident when confronting acoustic distribution shifts, encompassing both environmental degradation and speaker-inherent variations. For example, environmental factors, including reverberation~\cite{kinoshita2013reverb} and telephony bandlimiting~\cite{he10_interspeech},  dialectal and non-native speech \cite{lanehart2015language,thomas2015prosodic} affect both fine-grained and utterance-level acoustic characteristics, and physiological differences in child speech, such as shorter vocal tracts and higher fundamental frequencies~\cite{lee1999acoustics} can distort short-time spectral cues. Transformer-based encoders, which rely on global self-attention, often struggle to resolve localized acoustic phenomena, motivating adaptation strategies that explicitly model domain-specific context.

Although Transformer-based SFMs like Whisper are widely deployed, Conformer-based models \cite{gulati20_interspeech} achieve superior performance \cite{canary} on targeted ASR tasks \cite{open-asr-leaderboard}. As pretraining data distributions are opaque, it remains difficult to disentangle whether this gap is driven by the underlying training data or the inherent architectural superiority of the Conformer's local context modeling. Replacing deployed Transformer models is often impractical, raising the question of whether adaptation can bridge this architectural divide. Specifically, can a structurally "smarter" finetuning design imbue a Transformer with Conformer-like local acoustic properties?

To adapt speech foundation models (SFMs) to novel domains without the cost of full finetuning, Parameter Efficient Fine-Tuning (PEFT) methods~\cite{houlsby2019parameter}, which update only a small subset of parameters while freezing the backbone, have become popular. PEFT has also shown strong promise in speech adaptation \cite{fan2022draft}, with a key challenge being the modeling of local acoustic structure in addition to the long-range dependencies. To introduce locality, prior speech PEFT work has explored adapter-based designs~\cite{chen2023chapter,kheir2025parameter, wu24c_interspeech}, demonstrating the value of local inductive bias. In parallel, low-rank adaptation methods such as LoRA~\cite{hu2021lora} and its variants have become popular due to their minimal parameter overhead. LoRA has been applied to speech tasks including spoken language understanding~\cite{wang22y_interspeech}, accent-robust ASR~\cite{Bhatia2023}, child speech ASR~\cite{liu2023sparsely, wang2025ssvd, rolland2024exploring}, and multilingual ASR~\cite{xu2024towards, song2024lora}. Recent LoRA variants such as AdaLoRA~\cite{zhang2023adaptive} and DoRA~\cite{liu2024dora} improve linear low-rank adaptation through adaptive budget allocation and weight decomposition, respectively; in contrast, GC-LoRA changes the adapter inductive bias by adding gated local convolution. Concurrent rank/expert-based LoRA methods such as FlyLoRA~\cite{zou2026flylora} and Zipper-LoRA~\cite{mei2026zipperlora} address parameter interference through rank-level or expert-style decoupling, whereas GC-LoRA focuses on single-task acoustic adaptation. Closest to our work in ASR, \cite{kim24s_interspeech} adds convolution to projection-based PEFT bottlenecks, and Conv LoRA~\cite{zhong2024convolution, aleem2024convlora} explores convolutional low-rank updates in other domains. GC-LoRA differs by inserting a Conformer-style gated depthwise-separable convolution inside the LoRA bottleneck and targeting the attention output projection for local acoustic refinement.

This motivates the development of adapters that can inject local context modeling into frozen Transformer encoders. We thus propose Gated Convolutional LoRA (GC-LoRA), a structurally informed adapter modeled after the Conformer block. Different from prior locality aware PEFT for ASR that adds separate convolutional adapter blocks \cite{chen2023chapter}, GC-LoRA places a gated depthwise-separable convolution within the low-rank bottleneck so locality is learned through the same residual pathway used by LoRA. Evaluations on diverse, acoustically challenging datasets with a Whisper backbone demonstrate that GC-LoRA effectively bridges global Transformer processing and local acoustic modeling.

The main contributions of this work are summarized as:
\begin{itemize}
\item We introduce GC-LoRA, a parameter efficient adapter that integrates Conformer-style local convolution and gating mechanisms into pretrained Transformer encoders.
\item We show that GC-LoRA outperforms standard LoRA baselines across diverse domains (acoustically-degraded, bandlimited, dialectal, child), achieving notable Word Error Rate (WER) reductions with minimal computational overhead. \footnote{Our code and models can be found at \url{https://github.com/balaji1312/gc_lora}}
\end{itemize}

\section{Methodology}
\label{sec:methods}
\begin{figure}[t]
    \centering
    \includegraphics[width=0.84\columnwidth]{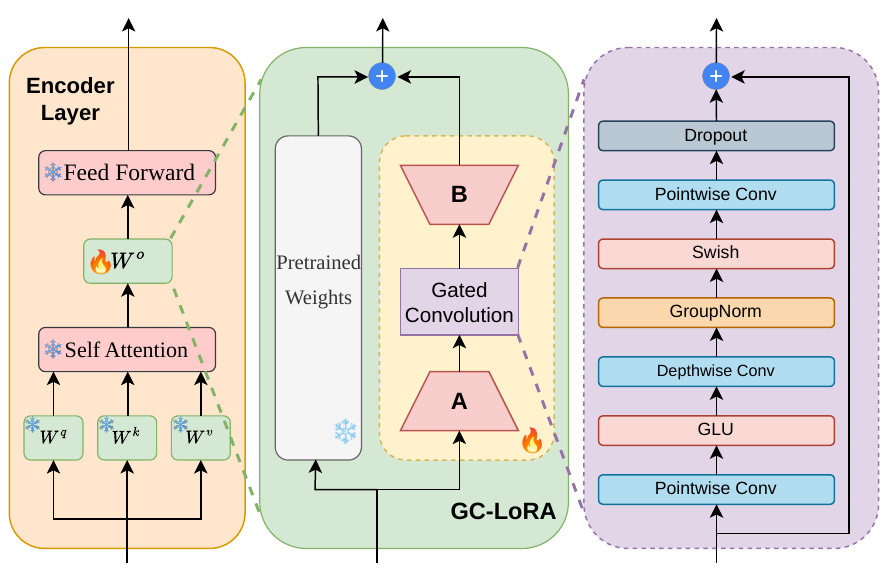}
    \caption{GC-LoRA adapter inserted into an encoder layer. Left: Baseline encoder layer. Middle: GC-LoRA replaces the standard residual low-rank update of $W_o$. Right: Gated Convolution block.}
    \label{fig:gclora_arch}
    \vspace{-0.1cm}
\end{figure}
We first review the Conformer architecture in Section \ref{sec:conformer} and LoRA in Section \ref{sec:lora}, then introduce the proposed Gated Convolutional LoRA (GC-LoRA) in Section \ref{sec:gc-lora}.

\subsection{The Conformer Block}
\label{sec:conformer}

While standard Transformers rely exclusively on multi-head self-attention (MHSA) to capture global sequence context, the Conformer architecture~\cite{gulati20_interspeech} demonstrates that the addition of a convolutional Module after the MHSA block to model local acoustic features significantly improves ASR performance. 

The module processes hidden states through a sequence of gating and depthwise convolutional operations. For an input sequence representation $\mathbf{X} \in \mathbb{R}^{S \times D}$ (where $S$ is the sequence length and $D$ is the embedding dimension), the module first applies a pointwise ($1 \times 1$) convolution along the temporal axis for channel expansion followed by a Gated Linear Unit (GLU) \cite{dauphin2017language}. The GLU acts as a dynamic feature-selection gate for local information. Following the gating mechanism, a 1D depthwise convolution is applied to capture temporal context, followed by normalization and a Swish activation \cite{ramachandran2017searching}. Finally, a second pointwise convolution projects the features, allowing the locally smoothed features to interact across channels. This explicit locality bias is largely absent in Transformer-only models motivating the need to inject these operations during finetuning.

\subsection{Low-Rank Adaptation (LoRA)}
\label{sec:lora}
Among PEFT methods, Low-Rank Adaptation (LoRA)~\cite{hu2021lora} has emerged as a widely used approach for adapting large Transformer models. LoRA freezes the pretrained weights and injects trainable low-rank matrices into selected layers, allowing efficient adaptation with minimal parameter updates. For a pretrained weight matrix $\mathbf{W} \in \mathbb{R}^{D \times D}$, LoRA models the output $\mathbf{H}$ as a low-rank decomposition:
\begin{equation}
    \mathbf{H} = \mathbf{X}(\mathbf{W} +\mathbf{\Delta W}) = \mathbf{X}\mathbf{W} + \frac{\alpha}{r} \mathbf{X} \mathbf{A} \mathbf{B},
\end{equation}
where $\mathbf{A} \in \mathbb{R}^{D \times r}$ and $\mathbf{B} \in \mathbb{R}^{r \times D}$ are trainable matrices and $\alpha$ is a scaling factor. LoRA is typically applied to the query and value projections of the multi-head self-attention layers \cite{hu2021lora}, enabling efficient recalibration of global attention patterns. However, its linear formulation lacks mechanisms to explicitly model local temporal context, motivating the proposed GC-LoRA architecture.

\subsection{Proposed: Gated Convolutional LoRA (GC-LoRA)}
\label{sec:gc-lora}

To enable frozen Transformer backbones to model local context without sacrificing parameter efficiency, we propose Gated Convolutional LoRA (GC-LoRA), which embeds the Conformer convolutional module inside the low-rank bottleneck of a parallel adapter. We specifically target the output projection matrix ($\mathbf{W}_o$) of the MHSA block to preserve standard global attention while allowing GC-LoRA to refine the attended features using local acoustic context, as illustrated in Figure \ref{fig:gclora_arch}. The GC-LoRA block thus acts as a residual path that smooths and refines the attended features based on their local acoustic context.

Given the input $\mathbf{X}$ to the output projection layer, GC-LoRA first compresses the features into a low-rank domain using the down-projection matrix $\mathbf{A}$:
\begin{equation}
    \mathbf{H}_{low} = \mathbf{X}\mathbf{A} 
\end{equation}
We apply the Pointwise Convolution expansion and Gated Linear Unit mechanism entirely within the low-rank space:
\begin{equation}
    \mathbf{H}_{glu} = \text{GLU}(\text{PointConv}(\mathbf{H}_{low})) 
\end{equation}
This gating enables selective modification of local context during adaptation. Next, we capture the local temporal neighborhood using a depthwise convolution, followed by Group Normalization \cite{wu2018group} and a Swish activation:
\begin{equation}
    \mathbf{H}_{dw} = \text{Swish}(\text{GroupNorm}(\text{DepthConv}(\mathbf{H}_{glu})))
\end{equation}
We substitute the standard Batch Normalization with Group Normalization, normalizing across channels for improved stability during Whisper finetuning with variable-length speech sequences and masked padding. A final pointwise convolution mixes the channels, and an internal residual connection is formed to anchor the learning process:
\begin{equation}
    \mathbf{H}_{res} = \mathbf{H}_{low} + \text{PointConv}(\mathbf{H}_{dw}) 
\end{equation}
Similar to LoRA, the tensor is then projected to the original embedding dimension using $\mathbf{B}$ and added to the output of the frozen pretrained weights:
\begin{equation}
    \mathbf{Y} = \mathbf{X}\mathbf{W}_o + \frac{\alpha}{r} (\mathbf{H}_{res} \mathbf{B})
\end{equation}

By isolating the convolution operations inside the $r$-dimensional bottleneck ($r \ll D$), GC-LoRA maintains a parameter footprint close to standard LoRA. We apply GC-LoRA to the attention output projection $W_o$: while $W_q$ and $W_v$ primarily shape attention patterns, $W_o$ receives the attended representation and is therefore a natural point to refine features with local acoustic context before the feed-forward layer. For the standard LoRA baseline, we follow prior work and apply linear LoRA to $W_q$ and $W_v$, where it is most effective~\cite{hu2021lora}. To isolate target-layer effects, we also compare against LoRA applied strictly to $W_o$ (LoRA-Output).
\section{Experiments}
\label{sec:experiments}

\subsection{Datasets}
\label{ssec:datasets}

\begin{itemize}

    \item \textbf{AMI~\cite{kraaij2005ami}:} The AMI Meeting Corpus consists of approximately 100 hours of multi-speaker meeting recordings captured with close-talking and far-field microphones, exhibiting overlapping speech, background noise, and reverberation. We include AMI to evaluate robustness to environmental acoustic degradation. We utilize the Kaldi AMI S5 recipe to obtain text-normalized transcripts and pre-segmented audio. Experiments are conducted on the Individual Headset Microphone (IHM) condition using the official train/dev/test splits.

    \item \textbf{Switchboard~\cite{godfrey1992switchboard}:} Switchboard-1 Release 2 is a corpus of approximately 260 hours of spontaneous conversational telephone speech from speakers across the USA. We include Switchboard to evaluate adaptation to frequency bandlimiting inherent to narrow-band telephony (8 kHz). Audio is upsampled to 16\,kHz for compatibility with Whisper. Data is segmented into utterance-level recordings and partitioned into train/validation/test splits with a 70/10/20 ratio.

    \item \textbf{CORAAL~\cite{kendall2023coraal}:} CORAAL is a corpus of sociolinguistic interviews featuring speakers of African American English (AAE). We include CORAAL to evaluate robustness to dialectal variation, as AAE exhibits phonological shifts and acoustic differences relative to Mainstream American English~\cite{johnson2024exploratory}. Following ~\cite{shankar2026compositional}, we train on the ATL, DTA, LES, DCA, DCB, and PRV regional subsets (137\,h), and hold out the ROC (13\,h) and VLD (12\,h) splits for development and testing. During preprocessing, recordings are segmented to retain only the interviewee's speech, with segments limited to 30\,s.

    \item \textbf{MyST~\cite{ward2011my}:} MyST is a corpus of conversational child speech from students in grades 3--5 interacting with a virtual science tutor. We include MyST to evaluate physiological acoustic shifts, as child speech exhibits different formants, higher fundamental frequencies, and greater articulatory variability than adult speech. Following~\cite{attia2023kid}, we use Whisper-large-v2 transcripts to remove utterances with WER~$>50\%$, fewer than three words, or duration~$>30\,\text{s}$, resulting in 133/21/25\,h for train/dev/test.

\end{itemize}

\subsection{Baselines and Ablations}
\label{ssec:baselines}
We compare GC-LoRA against established PEFT methods and structural ablations:

\begin{itemize}
    \item \textbf{Zero-Shot \& Full Finetuning:} We report the zero-shot and full model finetuning (updating all parameters) performance to serve as a reference for the frozen and the maximum parameter adaptation capacity of the network.
    \item \textbf{Standard LoRA~\cite{hu2021lora}:} Low-Rank Adaptation applied linearly to the Query ($\mathbf{W}_q$) and Value ($\mathbf{W}_v$) matrices of the MHSA blocks. This serves as our primary PEFT baseline.
    \item \textbf{LoRA-Output:} Low-Rank Adaptation applied linearly to the Output ($\mathbf{W}_o$) matrix of the MHSA blocks for a direct comparison with GC-LoRA.
    \item \textbf{Adapter~\cite{houlsby2019parameter}:} Bottleneck residual adapters inserted sequentially between the layers of the foundation model.
    \item \textbf{Conv-LoRA}~\cite{kim24s_interspeech,aleem2024convlora}: An ablation of our method where the Conformer-style gated depthwise-separable block is replaced by a single 1-D convolution inside the low-rank bottleneck. This isolates the contribution of GC-LoRA's gating and depthwise-separable design.
    \item \textbf{MultiConv-LoRA:} An extension of Conv-LoRA (similar in spirit to \cite{kheir2025parameter}) that employs multiple convolutional filters with varying kernel sizes ($k \in {7,15,23,31}$).
\end{itemize}

\subsection{Experimental Setup}
\label{ssec:setup}

Models are trained using the AdamW optimizer with a learning rate of $1\times 10^{-4}$, a linear decay schedule, a warmup ratio of 10\%, a batch size of 16, and trained for 10 epochs with early stopping based on validation WER. Low Rank Adapter (LoRA) modules are configured with rank $r=8$ and scaling factor $\alpha=16$, and residual adapters use a bottleneck dimension of 32. GC-LoRA utilizes a kernel size of $k=31$. All experiments are implemented using the Hugging Face Transformers library~\cite{wolf2020transformers}. Decoding is performed using greedy search, and the Whisper English text normalizer is applied prior to scoring. Statistical significance is computed using the NIST SCTK scoring toolkit \cite{NIST-SCTK}, employing a Matched-Pairs Sentence-Segment Word Error (MAPSSWE) test ($p < 0.05$). A comparison of different values for rank $r$ and kernel size $k$ for GC-LoRA is offered in Section \ref{ssec:ablations}. All experiments are conducted on a single NVIDIA RTX A4000 GPU. Inference profiling with batch size 1 shows negligible overhead relative to LoRA: GC-LoRA has identical peak VRAM, with latency/MACs increasing only from 57.6 ms/0.59G to 58.9 ms/0.63G.

\section{Results}
\label{sec:results}

\vspace{-5pt}
\begin{table}[!tb]
  \centering
  \setlength{\tabcolsep}{6pt}
  \renewcommand{\arraystretch}{1.1}
  \caption{Whisper-Medium adaptation results across four evaluation sets. We report WER on test set for AMI, Switchboard (SWBD), CORAAL and MyST. ``Params'' denotes number of parameters updated (Zero-shot updates none; Full FT updates the entire model). \textbf{Bold} indicates best results and $^*$ indicates statistical significance with $p <0.05$ among PEFT methods.}
  \label{tab:whisper_medium_main}
  \scriptsize
    \begin{tabular}{lrcccc}
      \toprule
      Method & Params & AMI & SWBD & CORAAL & MyST \\
      \midrule
      Zero-shot        & 0     & 16.4 & 17.2 & 17.0 & 13.1 \\
      Full FT          & 764M  & 10.8 &  5.7 &  9.8 &  8.9 \\
      \midrule
      LoRA \cite{hu2021lora} & 829k & 11.7 & 6.6 & 10.1 & 8.9 \\
      GC-LoRA (ours)   & 447k  & \textbf{11.5}\rlap{*} & \textbf{6.3}\rlap{*} & \textbf{9.9}\rlap{*} & \textbf{8.6}\rlap{*} \\
      \bottomrule
    \end{tabular}%
\end{table}

\begin{table}[!tb]
  \centering
  \scriptsize
  \setlength{\tabcolsep}{5pt}
  \renewcommand{\arraystretch}{1.1}
  \caption{Test-set WER (\%) comparison of LoRA and GC-LoRA across Whisper model sizes on AMI, Switchboard (SWBD), CORAAL, and MyST. ``L'' and ``GC'' denote LoRA and GC-LoRA, respectively.}
  \label{tab:lora_vs_conflora_scaling}
  \begin{tabular}{c|cc|cc|cc|cc}
    \toprule
    \multirow{2}{*}{Size}
      & \multicolumn{2}{c|}{AMI}
      & \multicolumn{2}{c|}{SWBD}
      & \multicolumn{2}{c|}{CORAAL}
      & \multicolumn{2}{c}{MyST} \\
    & L & GC & L & GC & L & GC & L & GC \\
    \midrule
    tiny     & 27.6 & \textbf{24.6}$^*$  & 18.8 & \textbf{18.1}$^*$  & 29.4 & \textbf{29.0}  & 15.9 & \textbf{15.7} \\
    base     & 19.3 & \textbf{17.8}$^*$  & 11.4 & \textbf{11.2}$^*$ & 20.3 & \textbf{20.1} & 12.7 & \textbf{12.2}\rlap{*}  \\
    small    & 14.5 & \textbf{13.4}$^*$  &  7.5 &  \textbf{7.4} & 12.7 & \textbf{12.3}\rlap{*}  &  9.8 & \textbf{9.6} \\
    medium   & 11.7 & \textbf{11.5}$^*$ &  6.6 &  \textbf{6.3}$^*$  & 10.1 &  \textbf{9.9}\rlap{*}  &  8.9 &  \textbf{8.6}\rlap{*}   \\
    large-v3 &  12.4    & \textbf{12.2}$^*$        &   7.4   & \textbf{7.0}$^*$     &     \textbf{9.7}  &  \textbf{9.7}    &  9.5 &  \textbf{8.7}\rlap{*}   \\
    \bottomrule
    
  \end{tabular}
  
\end{table}

\begin{table}[!tp]
  \centering
  \caption{Ablation Studies and Comparisons with other PEFTs. We report WER on test set for AMI, Switchboard (SWBD), CORAAL and MyST for Whisper-medium.}
  \label{tab:component_swb_myst_medium}
  \scriptsize
  \setlength{\tabcolsep}{6pt}
  \renewcommand{\arraystretch}{1.1}
  \begin{tabular}{lrcccc}
    \toprule
    Method & Params & AMI & SWBD & CORAAL & MyST \\
    \midrule
    GC-LoRA (ours) & 447k & 11.5 & \textbf{6.3} & \textbf{9.9} & \textbf{8.6} \\
    LoRA \cite{hu2021lora} & 829k & 11.7 & 6.6 & 10.1 & 8.9 \\
    \midrule
    LoRA-Output & 416k & 12.0 & 6.8 & \textbf{9.9} & 8.7 \\
    Adapter \cite{houlsby2019parameter} & 1.72M & \textbf{11.3} & 6.4 & 10.0 & \textbf{8.6} \\
    Conv-LoRA \cite{aleem2024convlora} & 1.75M & 11.9 & 6.5 & 10.2 & 8.7 \\
    MultiConv-LoRA & 1.77M & 11.7 & 6.5 & 10.1 & 8.7 \\
    \bottomrule
  \end{tabular}
\end{table}

\subsection{Effectiveness of GC-LoRA}
\label{ssec:core_result}
To  evaluate GC-LoRA across acoustic distribution shifts, we select four datasets: MyST (child), AMI (acoustically-degraded), CORAAL (dialectal), and Switchboard (SWBD) (narrowband). Table~\ref{tab:whisper_medium_main} reports Word Error Rate (WER) using the Whisper-medium backbone. Across all datasets, GC-LoRA consistently improves over the zero-shot baseline while updating only 447k parameters. Relative to standard encoder-only LoRA, GC-LoRA achieves statistically significant improvements ($p<0.05$) on all four test sets (AMI: 11.5 vs.\ 11.7; CORAAL: 9.9 vs.\ 10.1; SWBD: 6.3 vs.\ 6.6, MyST: 8.6 vs.\ 8.9) while using $\sim$46\% fewer trainable parameters (447k vs. 829k). Full finetuning provides the strongest overall performance, but at the cost of updating the entire 764M-parameter model. We note that on some datasets PEFT outperforms full finetuning; this phenomenon is consistent with prior observations in low-resource ASR, where high-capacity models overfit to limited training data without extensive regularization \cite{fan2024benchmarking}. The consistent gains across these varied domains, with a favorable accuracy--efficiency trade-off, demonstrate that injecting local context in GC-LoRA helps the Transformer bridge both environmental and speaker-related acoustic gaps.
\vspace{-8pt}
\subsection{Scalability Across Model Sizes}
\label{ssec:scalability}

Table~\ref{tab:lora_vs_conflora_scaling} compares LoRA (L) and GC-LoRA (GC) across Whisper model sizes. GC-LoRA yields consistent gains, with statistically significant improvements ($p{<}0.05$) in most settings. The largest relative improvement is observed on AMI with Whisper-tiny (27.6 to 24.6; 10.9\% relative). We note that on some datasets large-v3 exhibits slightly higher WER than medium. To keep comparisons controlled across model sizes, datasets, and PEFT variants, we fixed the adaptation hyperparameters and did not perform per-size tuning. Such non-monotonic scaling can arise in limited-data adaptation when higher-capacity models are more sensitive to fine-tuning hyperparameters and regularization \cite{fan2024benchmarking, ying25_wocci}. Overall, GC-LoRA yields consistent gains across model scales, particularly under acoustic mismatch.

\vspace{-8pt}
\subsection{Ablation Studies and Comparisons with other PEFTs}
\label{ssec:other_peft}

We investigate ablations of different components of GC-LoRA (Section~\ref{ssec:baselines}) as well as compare it with other PEFT techniques on Whisper-medium in Table~\ref{tab:component_swb_myst_medium}. GC-LoRA achieves the best WER on SWBD (6.3) and ties for the best performance on CORAAL (9.9) and MyST (8.6). While adapter-based methods attain the strongest AMI result (11.3), GC-LoRA remains competitive (11.5) and achieves these results with substantially fewer trainable parameters (447k vs.\ $\sim$1.7M). We also include LoRA variants LoRA-Output, Conv-LoRA, MultiConv-LoRA as ablations which replace the depthwise separable convolution. These methods are competitive but do not consistently outperform GC-LoRA, indicating that the combination of the $W_o$ target, gating, and depthwise-separable local convolution is important for the observed gains.
\vspace{-8pt}
\subsection{Impact of Rank and Kernel Size}
\label{ssec:ablations}
\begin{figure}[t]
    \centering
\includegraphics[height=0.2\textwidth,width=0.3\textwidth]{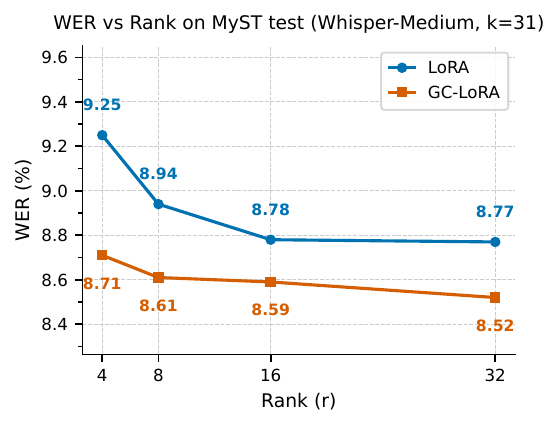}
    \caption{Impact of rank $r$ on MyST test WER for LoRA and GC-LoRA (Whisper-Medium; kernel size $k{=}31$).}
    \label{fig:size_impact}
\end{figure}
In Figure~\ref{fig:size_impact}, we analyze the effect of rank $r$ on MyST test WER (Whisper-Medium) by varying $r\in\{4,8,16,32\}$ for LoRA, and GC-LoRA with kernel size $k{=}31$. GC-LoRA remains stable across ranks and does not require tuning to achieve strong performance. Separately, we find GC-LoRA is largely insensitive to kernel size: with rank fixed to $r{=}8$, sweeping $k\in\{7,15,23,31\}$ changes WER by at most 0.13 (8.62--8.74), so we use $k{=}31$.

\subsection{Representation Analysis}
\label{ssec:representation}
\begin{figure}[t]
    \centering
    \includegraphics[width=0.75\columnwidth]{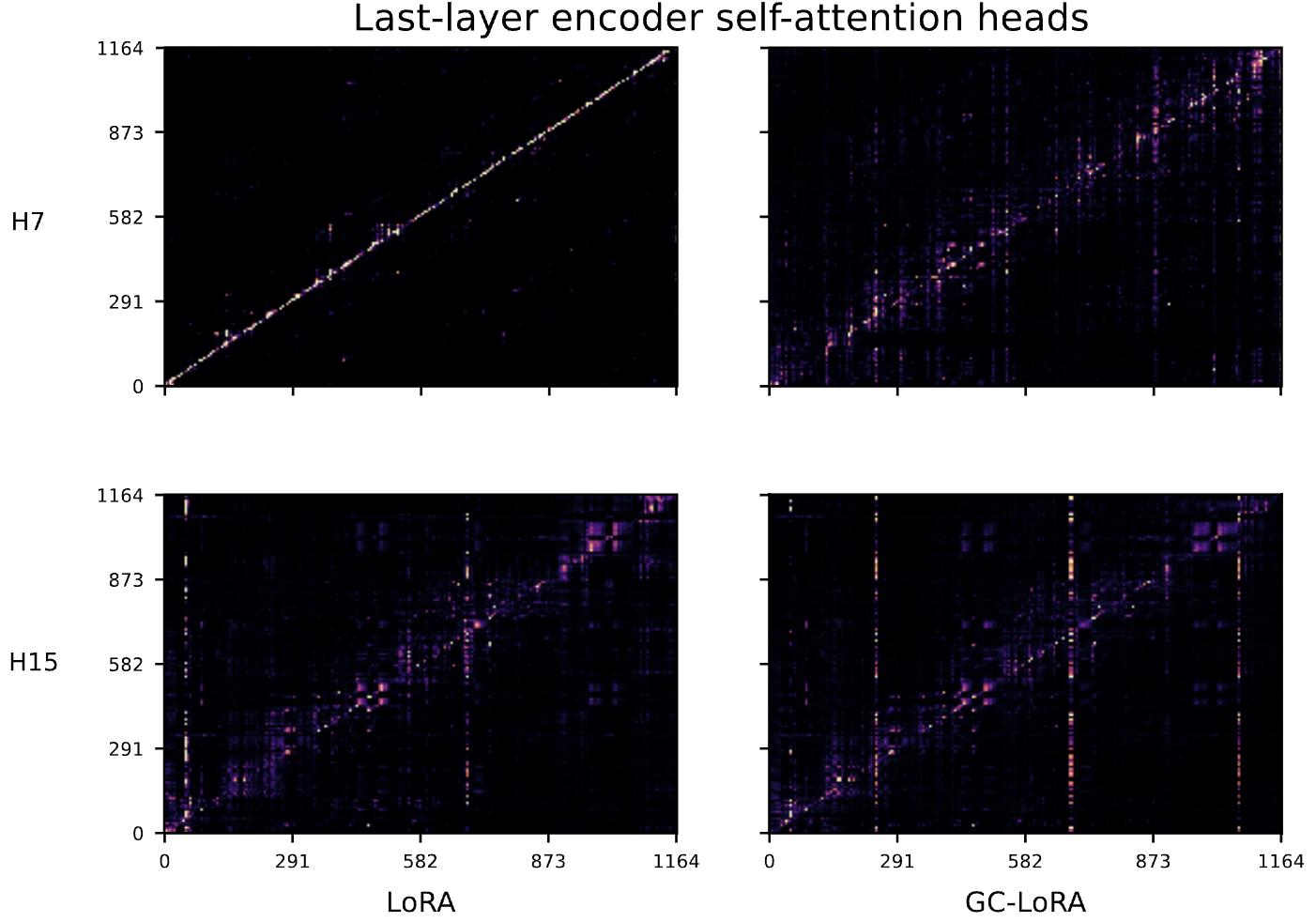}
    \caption{Attention map analysis of heads $h=7$ (top) and  $h=15$ (bottom) from the final encoder layer for a representative sample from the CORAAL test set: (a) LoRA and (b) GC-LoRA. The x-axis denotes encoder key position $j$, and the y-axis denotes encoder query position $i$; color intensity indicates the attention weight $A_{ij}$.}
    \label{fig:head_comp}
\end{figure}

Beyond ablations, we probe whether GC-LoRA's gains are consistent with altered locality in encoder representations. We analyze the final-layer encoder self-attention and quantify how far each head attends on average via the mean attention distance \cite{vig2019analyzing}:
\begin{equation}
\bar{d}=\frac{1}{NT}\sum_{h=1}^{N}\sum_{i=1}^{T}\sum_{j=1}^{T} |i-j| A^{(h)}_{ij}
\end{equation}
where $T$ is the encoder sequence length, $N$ the number of heads and  $A^{(h)}\in\mathbb{R}^{T\times T}$ the row-normalized attention map from head $h$. Fig.~\ref{fig:head_comp} visualizes attention maps for two final-layer heads with diagonal alignment \cite{yang20i_interspeech} for a representative utterance following prior ASR attention analyses \cite{peng2022branchformer}. For head $h{=}7$ GC-LoRA produces more diffuse attention pattern along the diagonal compared to LoRA. In head $h{=}15$, GC-LoRA exhibits more pronounced vertical bands, indicating that certain encoder positions attract attention from all query positions. To avoid cherry-picking, we report dataset-level $\bar{d}$ averaged over all utterances and heads. On CORAAL test, GC-LoRA yields a mean attention distance of 217.4 encoder tokens versus 213.8 for LoRA, a shift of 3.6 tokens. With GC-LoRA's short-range convolutional residual explicitly capturing local context in the output path, attention may be less required to focus strictly on nearby tokens, leading to slightly more diffuse attention while still improving ASR. This is consistent with observations in prior works \cite{shim2022understanding, prabhu24_interspeech}  where local context is handled by the convolutional branch and attention models broader dependencies. We emphasize this analysis is not diagnostic and that attention diffuseness alone does not guarantee improved recognition, but it supports our hypothesis for GC-LoRA's gains in acoustically diverse domains.
\vspace{-8pt}
\section{Conclusion}
\label{sec:conclusion}

In this work, we introduce GC-LoRA, a structurally informed Parameter Efficient Fine-Tuning approach designed to mitigate the vulnerability of Transformer-based Speech Foundation Models to acoustic distribution shifts. By embedding Conformer-style depthwise-separable convolutions and gating mechanisms directly into the attention output projections, GC-LoRA models local context without disrupting pretrained global representations. Evaluations across acoustic degradation, bandlimited speech, dialectal variations, and child speech demonstrate that GC-LoRA consistently outperforms standard LoRA despite requiring fewer trainable parameters, indicating that targeted local inductive biases are an efficient pathway to acoustic adaptation. Future work will explore extending this architecture to self-supervised backbones and acoustic encoders coupled with large language models.

\section{Acknowledgements}
\label{sec:ack}
This research is supported in part by the National Science Foundation (NSF) and the Institute of Education Sciences (IES), U.S. Department of Education (DoE), through Grant R305C240046 to the U. at Buffalo. The opinions expressed are those of the authors and do not represent views of the IES, DoE, or the NSF.

\section{Generative AI Use Disclosure}
\label{sec:genai_disclosure}
During the preparation of this work, the authors used ChatGPT (GPT-5.2 Thinking) for language editing, including proofreading and improving clarity and readability of the manuscript. All technical content, experimental design, results, and conclusions were produced and verified by the authors. After the use of Generative AI, the authors reviewed and edited the manuscript and take full responsibility for the content of the publication. Generative AI tools were not used to produce a significant portion of the manuscript and are not listed as authors.

\bibliographystyle{IEEEtran}
\bibliography{mybib}

\end{document}